\documentclass[12pt,preprint]{aastex}

\usepackage{psfig}

\def\cm{{\rm\,cm}}

\def\kms{{\rm\,km\,s^{-1}}}

\def\>{$>$}
\def\<{$<$}

\def\simlt{\lower.5ex\hbox{$\; \buildrel < \over \sim \;$}}
\def\simgt{\lower.5ex\hbox{$\; \buildrel > \over \sim \;$}}
\def\sqr#1#2{{\vcenter{\hrule height.#2pt
      \hbox{\vrule width.#2pt height#1pt \kern#1pt
         \vrule width.#2pt}
      \hrule height.#2pt}}}

\def\lsim{\lower.5ex\hbox{$\; \buildrel < \over \sim \;$}}
\def\gsim{\lower.5ex\hbox{$\; \buildrel > \over \sim \;$}}

\def\t{\ifmmode {\tau} \else $\tau$ \fi}

\def\ref{\noindent \hangafter=1 \hangindent=0.7 truecm}

\def\cm{\ifmmode {\rm cm}^{-1} \else cm$^{-1}$ \fi}
\def\s{\ifmmode {\rm s}^{-1} \else s$^{-1}$ \fi}
\def\cc{\ifmmode {\rm cm}^{-3} \else cm$^{-3}$ \fi}
\def\cs{\ifmmode {\rm cm}^{-2} \else cm$^{-2}$ \fi}
\def\g{\ifmmode \gamma \else $\gamma$\fi}
\def\G{\ifmmode \Gamma \else $\Gamma$\fi}

\def\kms{\ifmmode {\rm km\ s}^{-1} \else km s$^{-1}$\fi}

\begin{document}

\title{The ``Supercritical Pile" Model for GRB: Getting the 
$\nu F_{\nu}$ peak at 1 MeV}

\author{Demosthenes Kazanas, Markos Georganopoulos\altaffilmark{1}} 

\affil{NASA/GSFC, LHEA, Code 661, Greenbelt, MD, 20771}
\altaffiltext{1}{Also NAS/NRC Research Associate}


\author{Apostolos Mastichiadis}
\affil{Astronomy Department, University of Athens, Panepistimiopolis,
Athens, Greece}

\begin{abstract}
We propose that the internal energy of the GRB blast waves, thought to be
stored in the form of relativistic protons co-moving with the blast wave, 
is converted explosively (i.e. on light crossing time scales) into relativistic 
electrons of the same Lorentz factor, which are responsible for the 
production of observed prompt \g-ray 
emission of the burst. This conversion is the result of the combined effects 
of the reflection of photons produced within the flow  by upstream 
located matter, their re-interception by the blast wave and their eventual
conversion into $e^+e^--$pairs in interactions with the relativistic protons 
of the blast wave (via the $p \gamma \rightarrow e^+e^-$ reaction). This 
entire procedure is contingent on two conditions on the relativistic
protons: a kinematic one imposed by the threshold of the $p \gamma \rightarrow 
e^+e^-$ reaction and a dynamic one related to the column density of the 
post shock matter to the same process. 
This latter condition is in essence identical to that of the criticality 
of a nuclear pile, hence the terminology. It is argued that the properties of 
relativistic blast waves operating under these conditions are consistent with 
GRB phenomenology, including the recently found correlation between 
quiescence periods and subsequent flare fluence. Furthermore, it is shown 
that, when operating near threshold, the resulting GRB spectrum produces its 
peak luminosity at 
an energy (in the lab frame) $E \simeq m_ec^2$, thereby providing an answer to 
this outstanding question of GRBs.
\end{abstract}


\section{Introduction}\label{sect:intro}

The longstanding issue of the distance and absolute luminosity of GRBs
has been settled in the past decade through the observational evidence
collected  by $BATSE$ (Meegan et al. 1992) and $BeppoSAX$ (Costa et al.
1997), while the theoretical work of 
M\'esz\'aros \& Rees (1992) and Rees \& M\'esz\'aros (1992) provided the 
broader physical framework into which these events seem to generally fit. This 
framework associates GRBs with radiation emitted by relativistic blast
waves (hereafter RBWs), produced by an unspecified todate agent, presumably 
associated with the formation of a neutron star or black hole. While the 
source of the energy associated with GRBs has remained uncertain, there 
remains little doubt about the presence of the RBWs, which power also 
the later time emissions at X-ray, optical and radio frequencies, 
known collectively as GRB afterglows (see review of Piran 1999). 

With the discovery of GRB afterglows, much of the theoretical activity has 
since shifted to the study of the physics of these later time emissions.
Nonetheless, a number of issues associated with the prompt \g-ray 
emission, besides the nature of their energy source, still remain open.
Chief among them are the conversion of the RBW kinetic energy into 
radiation and the fact that the frequency at which the GRB luminosity 
peaks, $E_{\rm p}$, is narrowly distributed around a value intriguingly 
close to the electron rest mass energy. The purpose of the present 
note is to describe a process that  provides a ``natural" account of 
these generic, puzzling GRB features.

Following the work of Shemi \& Piran (1990) it has been 
generally accepted that a certain amount of baryons must 
be carried off with the blast waves responsible for the GRBs. This 
baryon contamination has even been deemed necessary, else the entire 
blast wave internal energy would be converted into radiation on very 
short time scales, leading to events of very different temporal and
spectral appearance (e.g. Paczy\'nski 1986) than observed in  GRBs . 
While the low radiative efficiency of baryons is essential for the
GRB energy transport to the requisite distances ($\gsim 10^{16}$ cm),
it becomes problematic when demanded that their
internal energy be radiated away on the short time scales associated 
with the GRB prompt \g-ray emission. Generally, this issue is 
sidestepped by appealing to an unknown process which transfers the 
proton energy into electrons (Dermer, B\"ottcher \& Chiang 1999), 
whose radiative evolution could then be accurately
computed.

The narrow range of the GRB $\nu F_{\nu}$ spectral peak energy, 
$E_{\rm p}$, is another well documented systematic feature of these 
events, a result of the extensive data base accumulated by $BATSE$. 
The compilation of Malozzi et al. (1995) shows clearly 
a preference for an energy $E_{\rm p} \simeq 200$ keV at 
which the $\nu F_{\nu}$ GRB spectra exhibit a maximum. In fact, 
it is precisely this maximum in the spectral energy 
distribution that  qualifies GRBs as such. 
Furthermore, when corrected for the redshift ($z_{\rm GRB} \sim 1$), 
$E_{\rm p}$ shifts close to the electron 
rest mass. While a compelling explanation of this fact is presently
lacking, several accounts have occasionally been proposed. For example, 
Brainerd (1994) argues that this is the result of down-Comptonization 
of a power law photon distribution that extends to $E \gg 1$ MeV by  
cold matter with Thompson depth $\tau_T \sim 10$, an explanation
possibly in conflict with the timing properties of GRBs (see 
e.g. Kazanas, Titarchuk and Hua 1997). The association of
the GRB emission with relativistically boosted synchrotron radiation from 
RBWs has made this particular issue far more acute, as the energy of
the latter should scale like $E_{\rm p} \propto \G^4$.
Therefore, even very small variations in the values 
of \G~ would lead to a very broad range in the values of $E_{\rm p}$.
Dermer et al. (1999) proposed that the observed distribution is the 
result of the time evolution of a blast wave with a specific baryon
loading, which when convolved with the triggering criteria of existing
detectors favors the detection of fireballs with $E_{\rm p}$ in the
observed range.
On the other hand, 
on the basis of analysis of SMM data,  Harris \& Share (1999) have 
argued that there is no apparent excess of GRBs with $E_{\rm p} 
\gg 1$ MeV, thus leaving this issue open. 

The present paper is structured as follows: In \S 2 we outline the 
fundamental notion behind our proposal for converting the RBW proton
energy into radiation, we derive the associated necessary conditions 
and discuss its relation to GRB phenomenology. 
In \S 3 we produce model spectra based on this 
proposal and indicate their relation to the particular value of 
$E_{\rm p}$ observed. Finally,
in \S 4 the results are discussed and certain conclusions are 
drawn.

\section{The ``Supercritical Pile"}\label{sect:pile}

The process described herein has been discussed in the 
past by Kazanas \& Mastichiadis (1999; hereafter KM99) in the 
context of AGN, where 
arguments have been put forward in favor of a hadronic origin of the 
relativistic electrons in blazars. This process is effectively the 
relativistic plasma instability proposed by Kirk \& Mastichiadis 
(1992; hereafter KM92) coupled to the increase in the photon energy 
associated with relativistically moving ``mirrors" (Ghisellini \& 
Madau 1997). While the mathematical formulation of the instability is 
given in detail in the above references, we provide below 
a qualitative re-derivation which focuses on and elucidates 
the underlying physics.

\subsection{A Static Plasma}\label{ssect:static}

Consider a spherical volume of size $R$, containing a relativistic proton 
plasma of differential spectrum $n_p(\g) = n_0 \g^{-\beta}$ (\g~ being 
the proton Lorentz factor), along with an (infinitesimal)
number of photons of energy $\epsilon$ (in units of $m_ec^2$); 
these photons can produce pairs via the $p\g \rightarrow e^-e^+$ 
reaction, provided that the proton population
extends to Lorentz factors 
$\g >\g_{\rm c}$ 
such that $\g_{\rm c} \, \epsilon \simeq2$. In the presence of a magnetic 
field $B$, the pairs (of Lorentz factor also equal to $\g_{\rm c}$)
produce synchrotron photons of energy $\epsilon_s = b \g_{\rm c}^2$
where $b = B/B_{\rm cr}$ is the magnetic field in units of the 
critical one $B_{\rm cr} = m_e^2 c^3/(e \hbar) \simeq 4.4 \, 10^{13}$ G.

For the reaction network to be {\sl self-contained} the energies of the seed
and synchrotron photons should be equal, yielding the kinematic
threshold of the process i.e. $\g_{\rm c} \epsilon_s =\g_{\rm c}^3 
\, b \simeq2$.
%
For the process to be also {\sl self-sustained}, at least one of the 
synchrotron photons must produce a pair before exiting the 
volume of the source. Hence, the optical depth of the source
to the $p\g \rightarrow e^-e^+$ reaction should be greater 
than $1/{\cal N}$, where ${\cal N} \simeq \g_{\rm c}/b \g_{\rm c}^2
= 1/b \g_{\rm c}$ is the total number of synchrotron photons produced by
an electron of energy $\g_{\rm c}$. This condition then reads
$\tau_{p\g} \simeq \sigma_{p\g} \, R \, n_p(\g)\g = \sigma_{p\g} 
\, R \, n_0 \g_{\rm c}^{-\beta+1} \gsim b \g_{\rm c}$. Eliminating
$\g_{\rm c}$ using the kinematic threshold condition $\g_{\rm c}^3 \, 
b \simeq2$, the critical column density expression reduces to 
\begin{equation}
 \sigma_{p\g} \, R \,  n_0 \gsim b^{1 - \beta/3}~,
\end{equation}
which (within factors of order unity) is the condition  
derived in KM92. This condition is 
similar to that of the criticality of a nuclear pile,
except that there one deals with neutrons rather than 
photons (this similarity carries also over in the case in the 
Comptonization of photons by hot electrons (Katz 1976)). 
It becomes apparent, hence, that the critical quantity
here (as also in a nuclear pile) is the column density (rather than 
the mass, the term ``critical mass" being simply a figure of speech). 
The pair - synchrotron photon - proton - pair network will be 
self-sustained if the column density is equal 
to the critical one. For larger values the number of pairs 
increases exponentially, eventually leading to a depletion of the
available energy source on time scales $\simeq R/c$ (a bomb!).

\subsection{Plasma in Relativistic Motion}\label{ssect:relativ}

KM99 extended the above analysis
to the case that the relativistic proton containing plasma moves
itself relativistically with Lorentz factor \G. 
The criticality conditions change quantitatively if
the radiation emitted by the plasma can be
``reflected" by matter located along its direction of motion. 
Due to relativistic beaming, essentially all the photons  produced
internally in the plasma will be focused in the forward direction, 
reflected and boosted in energy, upon their re-interception by 
the moving plasma, by a factor $\G^2$. If $b$ is 
the (normalized) value of the comoving magnetic field, the synchrotron
photons of energy $\epsilon_s = b \g^2$ will, upon their re-interception,
have energy $\epsilon_s = b \g^2 \G^2$, modifying the kinematic 
threshold condition to $b \g_{\rm c}^3 \G^2 \gsim 2$. 

The change in the kinematic threshold affects also the dynamic one: 
The photons necessary for the production of pairs in the comoving frame 
are now emitted by electrons of Lorentz factor only $\g_{\rm c}/\G$.  
The number of such photons is now ${\cal N} \simeq (\g/\G)/b (\g/\G)^2
=1/ b (\g/\G)$. Demanding again that the column of the plasma be 
greater than $1/{\cal N}$, along with the new threshold relation, 
leads to the condition
\begin{equation}
 \sigma_{p\g} \, \Delta_{\rm com} \,  n_{0 {\rm com}} \gsim b^{1 - \beta/3} \G^{-(1+2\beta/3)}
\end{equation}
namely the condition derived in KM99 ($\Delta_{\rm com}, \,  n_{0 {\rm com}}$
are the co-moving source size and density). Relativistic motion therefore, 
eases significantly the ``criticality"
condition in the case of a relativistically moving plasma.

The situation in the RBW of a GRB is somewhat different than that 
discussed just above. Though diffusive acceleration is very likely 
present in their associated shocks, the mere postshock isotropization 
of the flow creates, in the RBW frame, a relativistic population of protons
with mean energy $\langle E \rangle \simeq \G \, m_pc^2$. Therefore,
to be most conservative, one could dispense with the requirement of an 
accelerated proton population and demand that {\sl only} the protons of 
energy  $\G \, m_pc^2$ be present (these are certainly the most numerous). 
Therefore, upon setting $\g_{\rm c} 
\simeq \G$, the kinematic threshold condition reduces to $\G^5 \, b 
\gsim 2$. The number of photons emitted by electrons at threshold
then becomes ${\cal N} \simeq 1/b$ and the ``criticality" threshold
reads 
\begin{equation}
 \sigma_{p\g} \,\Delta_{\rm com} \,  n_{\rm com} = 
\sigma_{p\g} \, R \,  n \gsim b ~~~{\rm or} ~~~ \sigma_{p\g} 
\, R \, n \, \G^5 \gsim 2
\end{equation}
with the last relation incorporating both the kinematic and dynamic 
threshold ($\Delta_{\rm com} \,  n_ {\rm com}
= R \, n$ due to the Lorentz invariance of the column density. 
In this case, the quantity $n \, R$ simply denotes the 
amount of ambient matter per unit area swept by the RBW). 
For the typical values of $n$ and $R$ used in association with 
GRBs, i.e. $n = 1 \; n_0$ \cc and $R = 10^{16}\; R_{16}$ cm and considering
that $\sigma_{p\g} \simeq 5 \, 10^{-27}$ cm$^2$, the criticality 
condition yields $\G \gsim 180 \, (n_0 \, R_{16})^{-1/5}$, values well
within the accepted parameter range. 

Following the above analysis, the correlation between quiescence and
activity periods in GRBs found by Ramirez-Ruiz \& Merloni (2000),
finds a direct, qualitative interpretation: A RBW sweeps and `piles-up'
the ambient medium behind its forward shock, whose column density 
increases from below to above its critical value. Provided that 
the proper ``mirror" is located upstream, the internal energy, or part
of it depending on conditions, will be explosively converted into 
radiation. The energy released then would be proportional to the 
amount of matter accumulated during quiescence, with the process 
repeating as the blast wave progresses and more matter is accumulated.

\section{The Spectra}\label{sect:spectra}

Generally, the light curves as well as the GRB spectra are highly
variable. In the present model, the issue of variability becomes 
even more complex than in more conventional models, given that the 
injection rate of relativistic pairs depends on proton 
collisions with photons emitted at prior times, following their 
reflection by the ``mirror". 

While the computation of the spectra within this model is an inherently
time dependent problem, we present herein a simplified, steady-state 
treatment whose salient features, we believe, will be preserved 
in a more detailed calculation. 

We consider a RBW of energy $E = 10^{51} \, E_{51}$ erg 
in a uniform medium of density $n = 1 \, n_0$ \cc and of a (normalized)
opening half angle $\Theta = (\theta/ \pi)$. 
Because of the relativistic focusing of 
radiation,  we need only consider a section of the blast wave of 
opening half angle $\theta = 1/\G$. 
The shocked electrons of the ambient medium (and pairs from the 
$p \, \g \rightarrow e^+e^-$ process) produce synchrotron 
photons of energy $\epsilon_s \simeq b \, \G^2$. These,
upon their scattering by the ``mirror" and re-interception by the RBW,
are boosted to energy $\epsilon = \epsilon_s \, \G^2 = b \,
\G^4$ (in the RBW frame). These photons will then be scattered 
by (a) electrons of $\g \simeq 1$, originally contained 
in the RBW and/or cooled since the explosion, and (b) by the 
hot ($\g \simeq \G$), recently shocked ones to produce inverse Compton (IC)
radiation at energies correspondingly $\epsilon_1 \simeq b \, \G^4$ and 
$\epsilon_2 \simeq b \, \G^6$ at the RBW frame. 
At the lab frame, the energies of these three components, i.e. 
$\epsilon_s, \, \epsilon_1, \, \epsilon_2$ will be higher
by roughly a factor \G, i.e. they will be respectively at energies 
$b \, \G^3,~b \, \G^5$ and $b \, \G^7$. Assuming that the process operates 
near its kinematic threshold, $b\, \G^5 \simeq 2$, at the 
lab frame these components will be at energies $\epsilon_s \simeq \G^{-2}$,
$\epsilon_1 \simeq 2 \simeq 1$ MeV and $\epsilon_2 \simeq \G^2 \simeq$ 
10 GeV $(\G/100)^2$. This model therefore, produces 
``naturally" a component in the $\nu F_{\nu}$ spectral distribution
which peaks in the correct energy range. It also predicts the 
existence of another component at an energy $\G^2$ higher; such high
energy emission has been observed from several GRBs (Dingus 1995).

\begin{figure*}[t]
\centerline{\psfig{file=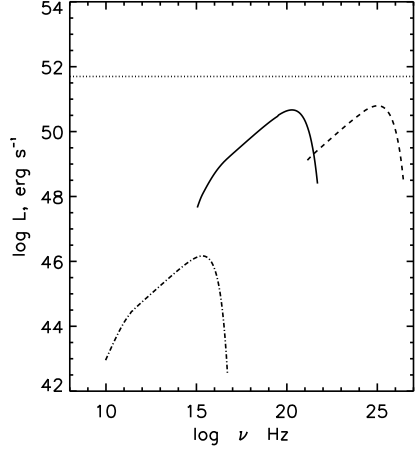,width=.4\textwidth,angle=0}}
\caption{The inferred isotropic luminosity of the three components 
discussed in the text for a burst with $\Delta \Omega /4 \pi = 0.005$,
\G = 240 and kinetic luminosity $L_{\rm k} \simeq 2 \times 10^{49}$ 
erg/s. The dot-dashed line is the synchrotron component, while the
solid and dashed lines are respectively the Compton scattered synchrotron 
by the cold and hot electrons. The dotted horizontal line denotes 
the equivalent isotropic kinetic luminosity. For a duration 
of $\simeq 20$ s. this will result in a fluence of $10^{52}$ ergs. }
\label{fig:temp}
\end{figure*}

The relative importance of these components depends on the specifics of
a given source. For the two IC components this depends on the 
ratio of the scattering depths of the cool $\tau_T(\g \simeq 1)$ and hot 
$\tau_T(\g \simeq \G)$ electrons. At a minimum, this ratio is equal 
to the ratio of their escape to cooling rates ${\cal R}_{\rm cool}/
{\cal R}_{\rm esc}\simeq t_{\rm esc}/t_{\rm cool}$. 
If $U_B = f_B (E/4\pi \, R^3 \, 
\Theta^2$) is the magnetic energy density ($f_B$  denotes 
departures from equipartition; $f_B \simeq 1$ in GRBs),  then
$ t_{\rm esc} \simeq \Delta_{\rm com}/c \simeq R/\G\, c$ and 
$t_{\rm cool} \simeq m_ec^2/(U_B \, \sigma_T \, c \, \G)$ so that
\begin{equation}
\frac{\tau_T(\g \simeq 1)}{\tau_T(\g \simeq \G)} =
\frac{f_B \, f_c}{\Theta^2 } \frac{E_{51}}{R_{16}^2} 
\end{equation}
where $f_c (>1)$ is a 
multiplicative factor indicating the contribution 
of cold electrons inherent in the RBW to the scattering depth. 
%
For $\Theta \sim 1/\G$ one obtains for the rate of photon scattering 
into the two IC components
\begin{equation}
\dot {\cal N}(E_\g\simeq \G^2) \simeq \frac{1}{f_B \, f_c \, \G^2} 
\frac{R_{16}^2}{E_{51}} \; \dot {\cal N}(E_\g \simeq 1)~,
\end{equation}
indicating that their luminosities would be roughly equal for the 
fiducial values of the parameters involved.

The relative luminosity between the direct synchrotron and the
cold-electron scattered synchrotron photons can be estimated
as follows: 
The number of electrons swept-up to radius $R$ by a section
of transverse dimension $d_t \simeq R/\G$ of the RBW is ${\cal N}_e \simeq 
n \, R^3 \, f_m/\G^2$, where $f_m (< m_p/m_e)$ is a multiplicative 
factor denoting the number of pairs produced in the RBW.  Given that 
each electron of energy \G~ produces $\simeq 1/b \G$ photons, 
the total number of photons produced to radius $R$ is
\begin{equation}
{\cal N}_\g \simeq \frac{n \, f_m}{b} \frac{R^3}{ \G^3} 
\end{equation}
These photons will be received in the lab frame at energy $\epsilon_s 
\simeq b \G^3$ over a time interval $\simeq R/ \G^2 c$ to yield 
a photon flux 
\begin{equation}
\dot {\cal N}_\g \simeq \frac{n \, f_m}{b} \frac{R^2 \, c}{ \G} 
\simeq n \, R \, \sigma_{\rm T} \, \G^4 \left( \frac{R \, c}{
\sigma_{\rm T}}\right) f_m
\label{synch}
\end{equation}
It is assumed that a fraction $\alpha$ of these photons will scatter 
at the ``mirror" and randomize producing 
a ``layer" of photons of width $\Delta \simeq R/\G^2$ (in the lab 
frame; it is assumed that the ``mirror" is thinner than $R/\G^2$), 
density $n_\g \simeq {\cal N}_\g \, \alpha /V = {\cal N}_\g 
\, \alpha /(R/\G)^2 \, \Delta$ and Thompson depth $\tau_{T,\g} \simeq
(n \, \sigma_{\rm T} R/b \, \G)\alpha \, f_m$ ($ \tau_{T,\g} \simeq n 
\, \sigma_{\rm T} \, R \, \G^4 \, \alpha \, f_m$ if near threshold 
i.e. for $b \, \G \simeq 1/\G^4$).

The number of electrons to radius $R$ swept by the section of the 
RBW considered here is then
${\cal N}_e \simeq (n R^3 \,/\G^2) (f_c+f_m)$;
hence the total number of photons scattered by the electrons
upon traversing the ``photon layer" of width $\Delta$ will be
${\cal N}_{sc,\g} \simeq {\cal N}_e \, \t_{T,\g}$. These will be 
received at the lab frame over a time $\simeq \Delta /c \G^2 
\simeq R /c \G^4$ to yield a photon flux (at energy $b \G^5 \simeq
2$)
\begin{equation}
\dot {\cal N}_{sc,\g} \simeq (n \, R \, \sigma_{\rm T})^2 \, \G^6 
\left( \frac{R \, c}{\sigma_{\rm T}}\right) \, \alpha \, f_m(f_c + f_m)
\label{ic1}
\end{equation}
suggesting that the ratio of the photon fluxes given by Eqs. (7), 
(8) will be proportional to $ n \, R \, \sigma_{\rm T} \, \G^2 
\alpha (f_c+f_m)$. For reasonable values of the source 
parameters one does obtain similar photon fluxes at these two energy 
bands ($b\G^3, b\G^5$) indicating that the luminosity at 
$\sim 1$ MeV is roughly $\G^2$ times that at the optical 
band. 

In figure 1 we present a sample of a spectrum obtained using
the arguments discussed above. Assuming the process to operate 
near threshold (with no accelerated proton component), leads to 
$\delta-$ function like injection of pairs; this results in the 
$\nu F_{\nu} \propto \nu^{1/2}$ spectrum of figure  1 (which
at low energies reverts to the $\nu F_{\nu} \propto \nu^{4/3}$
spectrum of thin synchrotron). 
Clearly more involved spectra will
result if particle acceleration is also incorporated in these
models; however, this will be the subject of future work.

\section{Discussion}

We have presented above a novel model for the prompt emission
of GRBs. We believe that this model provides some of the missing 
physics between the RBW proposal, which 
describes successfully the GRB energetics and time scales 
and the prompt emission of radiation (\g-ray as well as 
optical-UV), by a well defined mechanism for tapping the 
kinetic energy stored in the RBW baryonic component. 
Furthermore, the same physics employed in effecting the
conversion of baryon kinetic energy into radiation is
instrumental in producing a peak in the spectral energy
distribution at $E_{\rm p} \simeq 1$ MeV, thus providing a 
``natural" account of this GRB feature, an issue that 
has actually gotten even more puzzling with the advent of the 
RBW model for GRB. In addition
to this emission, the model implies the presence of \g-ray 
emission at higher energies, namely $E \simeq \G^2 m_ec^2$,
a fact supported by some observations already, but 
which will be explored in greater depth by SWIFT and GLAST, 
missions which will be able to test and put meaningful 
constraints on this specific model.

One of the fundamental features of this model is the presence
of the combination of kinematic and dynamic thresholds; this 
combination, along with the presence or absence of the 
necessary  ``mirror", make the model inherently time dependent.
At the same time, the kinematic threshold provides (for the 
first time to our knowledge) a regulating mechanism that puts
one of the peaks in the $\nu F_{\nu}$ distribution very close
to the observed value, despite the motion of the emitting plasma
with Lorentz factors of several hundreds.
Concerning the nature of the ``mirror" we are willing to speculate 
here that, because the main ``reflected" component consists of the 
prompt synchrotron photons which are emitted at optical frequencies,
one could use atomic cross-sections ($\simeq 10^{-16}~ {\rm cm}^2$) 
to estimate their reflected fraction $\alpha$; values of $n$, $R$
typical to those associated with GRBs yield $\alpha \simeq 0.01 - 1$.
This is a very rough estimate, because it ignores the ionization of 
the reflecting medium. A more detailed treatment 
including these effects is beyond the scope of the present work.

We would like to thank Jay Norris, Rob Preece and Brad Shaffer for stimulating
discussions and much information on GRB phenomenology.

{}

\end{document}